\documentclass[fleqn,usenatbib]{mnras}

\usepackage{newtxtext,newtxmath}
\usepackage[T1]{fontenc}

\DeclareRobustCommand{\VAN}[3]{#2}
\let\VANthebibliography\thebibliography
\def\thebibliography{\DeclareRobustCommand{\VAN}[3]{##3}\VANthebibliography}

\usepackage{graphicx}	% Including figure files
\usepackage{amsmath}	% Advanced maths commands
\usepackage{wasysym}    % Extra astronomy symbols
\usepackage{xspace}     % Auto-space at end of symbols.
\newcommand\Msun{\text{M}_{\astrosun}} % requires the wasysym package
\newcommand\Zsun{\text{Z}_{\astrosun}} % requires the wasysym package

\let\oldAA\AA
\renewcommand{\AA}{\text{\oldAA}\xspace}

\makeatletter % Workaround to only color the year for cite commands
  \patchcmd{\NAT@citex}
    {\@citea\NAT@hyper@{%
      \NAT@nmfmt{\NAT@nm}%
      \hyper@natlinkbreak{\NAT@aysep\NAT@spacechar}{\@citeb\@extra@b@citeb}%
      \NAT@date}}
    {\@citea\NAT@nmfmt{\NAT@nm}%
    \NAT@aysep\NAT@spacechar\NAT@hyper@{\NAT@date}}{}{}

  \patchcmd{\NAT@citex}
    {\@citea\NAT@hyper@{%
      \NAT@nmfmt{\NAT@nm}%
      \hyper@natlinkbreak{\NAT@spacechar\NAT@@open\if*#1*\else#1\NAT@spacechar\fi}%
        {\@citeb\@extra@b@citeb}%
      \NAT@date}}
    {\@citea\NAT@nmfmt{\NAT@nm}%
    \NAT@spacechar\NAT@@open\if*#1*\else#1\NAT@spacechar\fi\NAT@hyper@{\NAT@date}}
    {}{}
\makeatother

\title[The density-bounded twilight of starbursts]{The density-bounded twilight of starbursts in the early Universe}

\author[W. McClymont et al.]{William McClymont,$^{1,2}$\thanks{E-mail: \href{mailto:wjm50@cam.ac.uk}{wjm50@cam.ac.uk} (WM)}
Sandro Tacchella,$^{1,2}$
Francesco D'Eugenio,$^{1,2}$
Callum Witten,$^{3,1}$
Xihan Ji,$^{1,2}$
\newauthor
Aaron Smith,$^{4}$
Roberto Maiolino,$^{1,2,5}$
Santiago Arribas,$^{6}$
Jan Scholtz,$^{1,2}$
Charlotte Simmonds,$^{1,2}$
\newauthor
and Joris Witstok$^{1,2}$
\\%
\\%
$^{1}$Kavli Institute for Cosmology, University of Cambridge, Madingley Road, Cambridge CB3 0HA, UK\\%
$^{2}$Cavendish Laboratory, University of Cambridge, 19 JJ Thomson Avenue, Cambridge CB3 0HE, UK\\%
$^{3}$Institute of Astronomy, University of Cambridge, Madingley Road, Cambridge CB3 0HA, UK\\
$^{4}$Department of Physics, The University of Texas at Dallas, Richardson, Texas 75080, USA\\%
$^{5}$Department of Physics and Astronomy, University College London, Gower Street, London WC1E 6BT, UK\\%
$^{6}$Centro de Astrobiología (CAB), CSIC–INTA, Cra. de Ajalvir Km. 4, 28850- Torrejón de Ardoz, Madrid, Spain\\%
}

\date{Accepted XXX. Received YYY; in original form ZZZ}

\pubyear{2024}

\begin{document}
\label{firstpage}
\pagerange{\pageref{firstpage}--\pageref{lastpage}}
\maketitle

\begin{abstract}
The peculiar nebular emission displayed by galaxies in the early Universe presents a unique opportunity to gain insight into the regulation of star formation in extreme environments. We investigate 500 (109) galaxies with deep NIRSpec/PRISM observations from the JADES survey at $z>2$ ($z>5.3$), finding 52 (26) galaxies with Balmer line ratios more than $1\sigma$ inconsistent with Case B recombination. These anomalous Balmer emitters (ABEs) cannot be explained by dust attenuation, indicating a departure from Case B recombination. To address this discrepancy, we model density-bounded nebulae with the photoionisation code \texttt{CLOUDY}. Density-bounded nebulae show anomalous Balmer line ratios due to Lyman line pumping and a transition from the nebulae being optically thin to optically thick for Lyman lines with increasing cloud depth. The H$\alpha$/H$\beta$ versus H$\gamma$/H$\beta$ trend of density-bounded models is robust to changes in stellar age of the ionising source, gas density, and ionisation parameter; however, increasing the stellar metallicity drives a turnover in the trend. This is due to stronger stellar absorption features around Ly$\gamma$ reducing H$\beta$ fluorescence, allowing density-bounded models to account for all observed Balmer line ratios. ABEs show higher [\ion{O}{III}]/[\ion{O}{II}], have steeper ultra-violet slopes, are fainter, and are more preferentially Ly$\alpha$ emitters than galaxies which are consistent with Case B and little dust. These findings suggest that ABEs are galaxies that have become density bounded during extreme quenching events, representing a transient phase of $\sim$20\,Myr during a fast breathing mode of star formation.
\end{abstract}


%%%%%%%%%%%%%%%%%%%%%%%%%%%%%%%%%%%%%%%%%%%%%%%%%%

%%%%%%%%%%%%%%%%% BODY OF PAPER %%%%%%%%%%%%%%%%%%

\section{Introduction}
\label{sec:Introduction}

Galaxies exhibit remarkable variability throughout their evolution. While extensive studies have been carried out documenting their morphology \citep{Blanton:2009aa, Cappellari:2011aa}, kinematics \citep{Bundy:2015aa,Graham:2018aa}, and star formation \citep{Kennicutt:1983aa,Spindler:2018aa,McLeod:2021aa} in the nearby Universe, the early evolution of galaxies remains at the frontier of observational and theoretical work. Nebular emission lines, having already been used to study fundamental galaxy properties locally, such as gas-phase metallicities \citep{Maiolino:2019aa} and electron densities \citep{Kewley:2019ab}, provide a crucial window into the interstellar medium (ISM) of these juvenile galaxies. Studies of nebular emission have revealed that galaxies in the early Universe hosted extreme episodes of star formation \citep{Tacchella:2023aa, Endsley:2023aa}, rapid quenching \citep{Looser:2023ab, DEugenio:2024ab}, and luminous active galactic nuclei \citep[AGN;][]{Harikane:2023aa,Maiolino:2024ac}.

Interpreting and modelling of nebular emission in galaxies with a complex hierarchy of ISM substructure is fraught with difficulty, and so approximations are used to reduce the complexity. One of the most common assumptions that is made involves the treatment of recombination cascades of ions, which are typically not solved explicitly due to computational constraints. In fact, only a few codes solve for the level populations in sufficient detail, and those that do must compromise elsewhere their modelling (e.g. limited geometry in \citealt{Ferland:2017aa}). The predominant approximation in galactic environments is to assume Case B recombination, which is one of four approximations known as Case A, B, C, and D (see \citealp{Chakraborty:2021aa} for a more detailed discussion). Case A recombination is valid where gas is optically thin to Lyman line emission \citep{Menzel:1937aa,Baker:1938aa}. In this case, all photons emitted as a result of recombination cascades escape the gas, including any ionising photons arising from recombination of electrons to the ground state. Case B is generally the most reasonable approximation for galaxies, and it is valid where the gas is optically thick to Lyman series photons \citep{Menzel:1937ab,Baker:1938aa}. Lyman line photons are reabsorbed and converted to other line photons, with the exception of Ly$\alpha$, which must eventually escape, be converted to two-photon emission, or be destroyed by dust. Additionally, recombinations which produce a free-bound continuum photon with sufficient energy are assumed to immediately ionise a nearby hydrogen atom. Cases C and D are analogous to Cases A and B respectively, except that the fluorescence of lines by continuum photons is accounted for \citep{Baker:1938ab,Ferland:1998aa,Luridiana:2009aa}.

Case B recombination predicts specific values of the intrinsic Balmer line ratios, for example, H$\gamma$/H$\beta=0.47$ and H$\alpha$/H$\beta=2.86$ for gas at 10\,000\,K and with an electron density of $10^2~\mathrm{cm^{-3}}$. These ratios can be changed somewhat by the temperature of the gas and the electron density. However, for reasonable gas temperatures, the range of Balmer line ratios possible under Case B is narrow, with H$\alpha$/H$\beta$ changing by only $\sim$4\% from 10\,000\,K to 20\,000\,K \citep{Osterbrock:2006aa}. Variations of electron density has a similarly small effect, with H$\alpha$/H$\beta$ varying by 2\% between $10^2~\mathrm{cm^{-3}}$ and $10^6~\mathrm{cm^{-3}}$ for gas at 10\,000\,K. This relative stability is exploited extensively, particularly when correcting for dust effects in observed spectra \citep[e.g.][]{Oh:2011aa,Maheson:2024aa}. Any significant departures from Case B could lead to misinterpreting the observationally inferred properties of galaxies, and so it is important to understand the limitations of this assumption in the early Universe. 

Dwarf and high-redshift galaxies are typically characterised by strong bursts of star formation, which has been studied extensively in numerical work \citep[e.g.][]{Hopkins:2014aa,Hayward:2017aa,McClymont:2025aa} and observationally \citep[e.g.][]{Looser:2023ab, Faisst:2024aa,Simmonds:2024aa}. Such bursty star-formation histories have been invoked to explain a wide range of observations, such as the presence of dark matter cores in dwarf galaxies \citep{Read:2005aa,Pontzen:2012aa} and the ultra-violet (UV) luminosity function at high redshifts \citep{Ren:2019aa,Sun:2023aa}. Theoretical models predict that the star formation rate (SFR) stochasticity should increase with redshift \citep{Faucher-Giguere:2018aa,Tacchella:2020aa,Kravtsov:2024aa,McClymont:2025aa}, which is now supported by observations of objects undergoing extreme starbursts at high-redshift \citep{Endsley:2023aa,Dressler:2024aa} and which have recently been rapidly quenched, or mini-quenched \citep{Looser:2024aa,Trussler:2025aa}. Such violent burst and quenching episodes are expected to have dramatic effects on their gas, including driving high-mass outflows \citep{Hayward:2017aa,Hopkins:2023aa}. If the extreme environments of galaxies undergoing burst-quench cycles cause a breakdown in the validity of Case B assumptions, then identifying and studying such galaxies would provide a unique opportunity to gain further insights into these processes.

Recent studies have revealed significant deviations from expected Case B recombination Balmer line ratios. \citet{Yanagisawa:2024ac} analyse GS9422 \citep[$z=5.9$,][]{Cameron:2023ab} and RXCJ2248-ID \citep[$z=7.1$,][]{Topping:2024aa}, which have lower H$\alpha$/H$\beta$ and higher H$\gamma$/H$\beta$ values than Case B. Dust attenuation would act to increase H$\alpha$/H$\beta$ and decrease H$\gamma$/H$\beta$, and therefore cannot account for the Balmer line emission in these galaxies. This indicates that there is a genuine intrinsic departure from Case B. Anomalous Balmer emitters (ABEs) --- galaxies that exhibit Balmer line ratios inconsistent with Case B --- have also been investigated at lower redshift. \citet{Scarlata:2024aa} place a particular focus on SXDF308 \citep[$z=0.0695$,][]{Mehta:2018aa}, however they also find evidence for a wider population of ABEs using data from the SDSS survey. Observations of ABEs also extend to Ly$\alpha$ emitters at $z\sim0.3$ (\citealt{Atek:2009aa}) and galaxies from $1<z<3.5$ (\citealt{Pirzkal:2024aa}), suggesting that the conditions for departure from Case B manifest more broadly and deserve greater attention in astrophysical studies.

Both \citet{Yanagisawa:2024ac} and \citet{Scarlata:2024aa} have the same preferred explanation for ABEs; they claim that the assumption that these galaxies are optically thin to Balmer lines has broken down. \citet{Yanagisawa:2024ac} propose that if emission nebulae are surrounded by a shell of gas which is optically thick to Balmer lines, this would shift the line ratios to higher H$\gamma$/H$\beta$ and lower H$\alpha$/H$\beta$, the apparent trend of GS9422 and RXCJ2248-ID. However, this scenario would require a relatively high column density of \ion{H}{I} in the excited $\mathrm{n=2}$ state, and the physical origin of this excited gas remains unclear. \citet{Yanagisawa:2024ac} also consider density-bounded nebulae, which are nebulae where ionising photons leak from a cloud of gas, as a possible explanation. Density-bounded nebulae can cause anomalous Balmer line ratios because their Balmer line ratios vary as a function of depth in a cloud, only aligning with Case B values when the cloud becomes sufficiently optically thick to Lyman lines. However, they conclude that this solution is less preferred because it requires fine tuning of nebular cloud depth to explain the line ratios of GS9422 and RXCJ2248-ID. On the other hand, \citet{Scarlata:2024aa} explicitly consider Case C, which includes the continuum pumping of Lyman lines. However, they ultimately dismiss density-bounded nebulae as an explanation for the anomalous Balmer line ratios in SXDF308 because they cannot find models that match the observed H$\alpha$/H$\beta$ and H$\gamma$/H$\beta$ while also reproducing [\ion{O}{III}]/[\ion{O}{II}] and the H$\beta$ equivalent width.

Still, an important ingredient in line emission modelling is the spectral energy distribution (SED) of the ionisation source. \citet{Yanagisawa:2024ac} employ black-body spectra, which can lead to unrealistic continuum pumping of emission lines when compared to stellar SEDs, and drastically impact the predicted evolution of Balmer line ratios as a function of cloud depth. While \citet{Scarlata:2024aa} do utilise stellar SEDs, their models do not account for variations in stellar metallicity. This means that they were not able to consider the varying Lyman line absorption spectra of stars with differing metallicities. In this paper, we address these limitations by constructing realistic density-bounded nebulae models that incorporate stellar SEDs with a range of metallicities.

In order to constrain different non-Case B scenarios, we must identify more galaxies with anomalous Balmer line ratios, as GS9422 and RXCJ2248-ID cannot paint the full picture alone. With this goal in mind, we use the publicly available spectra and flux catalogues from the JADES survey covering the GOODS-N and GOODS-S fields \citep{Bunker:2024aa, DEugenio:2025aa}. This dataset is particularly valuable for its extensive spectral coverage, which allows us to explore the H$\alpha$/H$\beta$ ratios of galaxies at $z>2$. We specifically focus on galaxies at $z>5.3$ in order to reliably decompose H$\gamma$ and [\ion{O}{III}]$\lambda$4363~\AA\ and therefore compare the H$\alpha$/H$\beta$ versus H$\gamma$/H$\beta$ trends predicted by our models against the observations.

The paper is structured as follows. In Section~\ref{sec:Observations}, we present our observed sample of galaxies from the JADES survey. In Section~\ref{sec:Anomalous Balmer line ratios}, we show the H$\alpha$/H$\beta$ and H$\gamma$/H$\beta$ ratios observed in these galaxies, focusing on ABEs. We present our models of density-bounded nebulae and compare the Balmer line ratios produced by these models to the observed data.
In Section~\ref{sec:Other spectral features}, we explore the continuum features, Ly$\alpha$ emission, [\ion{O}{III}]/[\ion{O}{II}], and [\ion{O}{III}]/H$\beta$ of ABEs and discuss how they relate to density-bounded nebulae. 
In Section~\ref{sec:Discussion}, we discuss the broader implications of our results on our understanding of the ISM of galaxies in the early Universe.
We conclude in Section~\ref{sec:Conclusions}, where present our main results.

Throughout this paper, we will employ the following shorthand notations for convenience: [\ion{O}{II}] refers to [\ion{O}{II}]$\lambda$3726,3729~\AA, and [\ion{O}{III}] denotes [\ion{O}{III}]$\lambda$4959,5007~\AA. We use the AB magnitude system and assume the Planck15 flat $\Lambda$CDM cosmology \citep{Planck-Collaboration:2016aa} with $\Omega_m=0.308$ and $H_0=67.8$ km/s/Mpc, using the Astropy package to perform cosmology calculations \citep{Astropy-Collaboration:2013aa,Astropy-Collaboration:2018aa,Astropy-Collaboration:2022aa}. 

\section{Observations}
\label{sec:Observations}

The high-redshift galaxy sample in this work was observed with \textit{JWST} NIRSpec as a part of the \textit{JWST} Advanced Deep Extragalactic Survey \citetext{JADES, \citealp{Eisenstein:2023aa}; PID 1180 and~1181, PI: D. Eisenstein; PIDs: 1210 and~1286, PI: N. L\"utzgendorf.  \citealp{Bunker:2023aa,DEugenio:2025aa}}, later extended with parallel NIRSpec observations of the JADES Origins Field \citetext{PID 3215, PIs: D.~Eisenstein and R.~Maiolino; \citealp{Eisenstein:2023ab}}. The emission line fluxes and spectra are sourced from the third JADES public release data, presented in \citet{DEugenio:2025aa}. This dataset provides fully reduced spectra, along with an emission line catalogue for lines detected with a signal-to-noise ratio $S/N>5$.
The background subtraction and slit-loss corrections are optimised for point-like sources, appropriate for compact, high-redshift galaxies. We exclusively use PRISM catalogue spectra and emission line fluxes in this work, which were obtained using the \textsc{ppxf} spectral fitting code \citep{Cappellari:2023aa}.

We consider two samples in this work. Firstly, we select galaxies at $z>2$ with robust 5$\sigma$ detections of both H$\alpha$ and H$\beta$ in order to look for sub-Case B H$\alpha$/H$\beta$ across a large sample. There are 500 galaxies in this sample, 109 of which are at $z>5.3$. Secondly, our more focused sample consists of galaxies at $z>5.3$ with 5$\sigma$ detections of both H$\alpha$, H$\beta$, and H$\gamma$. This redshift threshold is dictated by the wavelength-dependent spectral resolution of the NIRSpec prism \citep{Jakobsen:2022aa}, which increases with wavelength.
At $z>5.3$, the effective prism spectral resolution is $R\approx 120$, as determined from the nominal $R$ reported in \citet{Jakobsen:2022aa}, increased by an average 40~percent to take into account the typical size of the JADES targets, which underfill the MSA shutters giving an increase in spectral resolution above the nominal value. With $R=120$, we obtain a $\sigma=15~\AA$, sufficient to resolve the H$\gamma$--[\ion{O}{III}]$\lambda$4363~\AA complex. This sample contains 39 galaxies and it is crucial because the H$\alpha$/H$\beta$ versus H$\gamma$/H$\beta$ trends can allow us to discriminate between different non-Case B physics. Sub-samples taken from these parent samples are defined when relevant (e.g. Section~\ref{sec:Continuum features and Lyman-alpha}).

\section{Anomalous Balmer line ratios}
\label{sec:Anomalous Balmer line ratios}

In this section we detail Anomalous Balmer emitters (ABEs) observed in the early Universe. In Section~\ref{sec:Balmer line observations} we explore H$\alpha$/H$\beta$ and H$\gamma$/H$\beta$, and demonstrate that many galaxies appear to have anomalous Balmer line ratios. In Section~\ref{sec:Density-bounded nebulae models} we introduce our modelling approach to density-bounded nebulae. In Section~\ref{sec:Comparison to observed line ratios} we compare our density-bounded models to observed H$\alpha$/H$\beta$ and H$\gamma$/H$\beta$ ratios. In Section~\ref{sec:Anatomy of density-bounded nebulae} we explore the physical origin of the line ratio trends.

\subsection{Balmer line observations}
\label{sec:Balmer line observations} 

\begin{figure}
    \centering
	\includegraphics[width=\columnwidth]{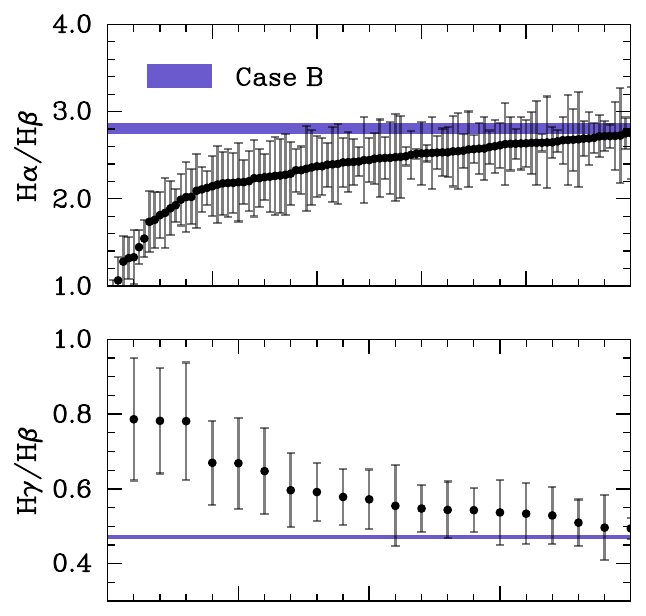}
    \caption{Balmer line ratios for $z>2$ galaxies compared to Case B values in the 10\,000\,--\,20\,000\,K temperature range (blue shaded region). \textit{Top:} H$\alpha$/H$\beta$ for galaxies  with 5$\sigma$ detections of H$\alpha$ and H$\beta$, ordered by increasing line ratio. The 100 galaxies with the lowest H$\alpha$/H$\beta$ ratios are shown. 52 (14) galaxies with $z>2$ ($z>5.3$) show Balmer decrements which are more than $1\sigma$ below the Case B values, which cannot be explained by dust attenuation. \textit{Bottom:} H$\gamma$/H$\beta$ values for galaxies which also have a 5$\sigma$ detection of H$\gamma$, ordered by decreasing line ratio. The 20 galaxies with the highest H$\gamma$/H$\beta$ ratios are shown. 12 galaxies with $z>5.3$ show values more than $1\sigma$ above the Case B range, which cannot be due to dust attenuation.}
    \label{fig:Ha_Hb_caseB}
\end{figure}

For our initial investigation, we select all galaxies at $z>2$ with robust $5\sigma$ detections of H$\alpha$ and H$\beta$. We show the H$\alpha$/H$\beta$ values of these galaxies in the top panel of Fig.~\ref{fig:Ha_Hb_caseB}, and it is clear that a significant fraction of these galaxies have H$\alpha$/H$\beta$ ratios that are inconsistent with Case B and not accessible through dust attenuation. 52 of the 500 galaxies have a H$\alpha$/H$\beta$ which is at least 1$\sigma$ below Case B\footnote{We also fit the H$\alpha$ and H$\beta$ lines ourselves with a simple Gaussian fitting procedure for all ABEs and found similar anomalous line ratios as in the JADES catalogue. However, for the few most extreme examples (H$\alpha$/H$\beta<2$) in the emission line catalogue, we found values of H$\alpha$/H$\beta\approx2$. We retain the catalogue values for the sake of consistency, but note that the most extreme deviations from Case B (H$\alpha$/H$\beta<2$) may be overestimated.}. 14 of those galaxies are at $z>5.3$, out of the total 109 galaxies in the sample at $z>5.3$. This indicates that there is indeed a population of galaxies in our sample for which Case B does not hold. Given that even a relatively small amount of dust attenuation could increase an anomalous H$\alpha$/H$\beta$ value enough such that it appears to agree with Case B, we can only identify ABEs with H$\alpha$/H$\beta$ when they have little dust.

In the bottom panel of Fig.~\ref{fig:Ha_Hb_caseB} we show the H$\gamma$/H$\beta$ values for galaxies at $z>5.3$ which have a 5$\sigma$ detection of H$\gamma$ and H$\beta$. 12 of the 39 galaxies show H$\gamma$/H$\beta$ which is at least 1$\sigma$ above Case B. Of these 12 galaxies, 10 have H$\alpha$/H$\beta$ ratios consistent with Case B. This may be due to dust attenuation having a stronger effect on H$\alpha$/H$\beta$ compared to H$\gamma$/H$\beta$, and indicates that some galaxies that appear to be consistent with Case B may, in fact, be anomalous galaxies with dust attenuation. The H$\alpha$ flux is blended with [\ion{N}{II}]$\lambda$6585~\AA, and is therefore an upper bound, which means that some galaxies that appear to be in agreement with Case B may not be if we could resolve H$\alpha$ alone. However, we do not expect this effect to be important. [\ion{N}{II}]$\lambda$6585~\AA\,is unlikely to be significant at the low metallicity values observed in galaxies at these redshifts \citep{Simmonds:2023aa, Cameron:2023aa}. \citet{Sandles:2024aa} confirmed the small contribution of [\ion{N}{II}] by performing a stack of all galaxies at $z>4$ with medium resolution spectra, and finding H$\alpha$/(H$\alpha$+[\ion{N}{II}])=0.96.

\begin{figure}
    \centering
	\includegraphics[width=\columnwidth]{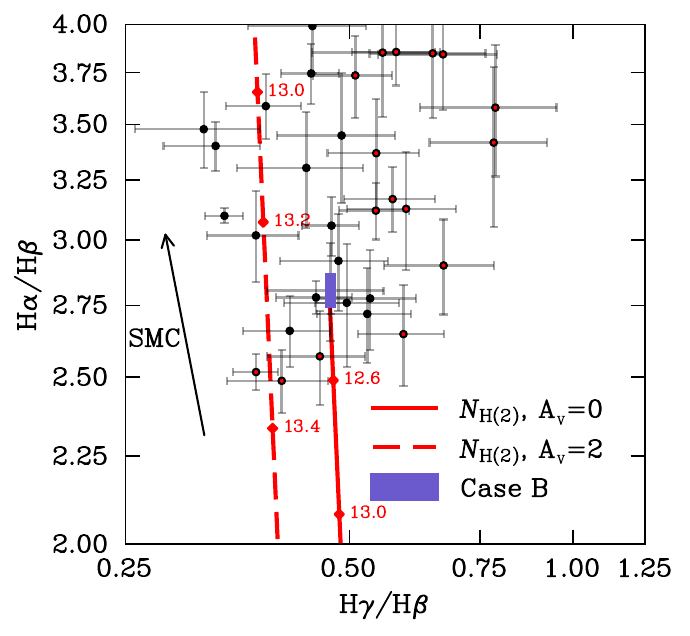}
    \caption{H$\alpha$/H$\beta$ against H$\gamma$/H$\beta$ for galaxies with 5$\sigma$ detections of H$\alpha$, H$\beta$, and H$\gamma$. The arrow represents increasing dust attenuation for an SMC extinction curve ($\mathrm{R}_{\rm _V}=2.7$) with a length of $\mathrm{A}_{\rm _V}=1$ mag \citep{Gordon:2003aa}. A \citet{Cardelli:1989aa} dust curve with $\mathrm{R}_{\rm _V}=3.1$ (not shown) has a nearly identical direction. The blue box around H$\alpha$/H$\beta=2.8$ and H$\gamma$/H$\beta=0.47$ represents Case B values in the 10\,000\,--\,20\,000\,K temperature range. Many galaxies are not consistent with Case B line ratios and dust attenuation; galaxies which are at least 1$\sigma$ inconsistent are marked with red points. Following \citet{Yanagisawa:2024ac}, the solid (dashed) red line represents Balmer absorption due to a shell of HI gas in the $n=2$ state, combined with dust $\mathrm{A}_{\rm _V}=0$ mag ($\mathrm{A}_{\rm _V}=2$ mag). The log column density of the excited neutral gas is labelled on the lines. This model of an excited hydrogen shell cannot explain galaxies with elevated H$\gamma$/H$\beta$.}
    \label{fig:Ha_Hg_vs_Ha_Hb_caseB}
\end{figure}

To further test the statistical significance of Case B deviation in our sample, we perform Monte Carlo simulations of the measurement errors. Our null hypothesis is that all intrinsic ratios are consistent with Case B, so we assume that all Balmer line ratios which are inconsistent with Case B are due to measurement errors, and therefore set them to the Case B value. In order to be as conservative as possible, we choose the Case B values at 20\,000\,K and $10^6~\mathrm{cm^{-3}}$ (H$\alpha$/H$\beta=2.725$ and H$\gamma$/H$\beta=0.476$). We sample 1,000,000 times from the error around each data point, and calculate the probability that the observed number of apparently Case B violating galaxies can be reproduced by a population of Case B galaxies and measurement errors. We find p-values of 0.012 and 0.012 for the H$\alpha$/H$\beta$ and H$\gamma$/H$\beta$ samples, respectively, indicating that the violation of both line ratios is statistically significant for our sample of galaxies. We note that using the less conservative and usually adopted Case B values at 10\,000\,K, we find p-values of 0.0011 and 0.00041 for the H$\alpha$/H$\beta$ and H$\gamma$/H$\beta$ samples, respectively.

The statistical tests thus far simply quantify whether the number of Case B violating galaxies can be successfully reproduced. As a further test we also calculate the probability that a Case B population could produce a population of galaxies with average offsets from the Case B ratio which is at least as extreme as the offsets for our observed Case B-violating sample. We again use the more conservative Case B values at 20\,000\,K. Also in the interest of producing a conservative estimate, we remove galaxies with the most extreme ratios, H$\alpha$/H$\beta<1.6$, where the line ratio measurements may be less reliable. We recover p-values of 0.00049 and 0.00003 H$\alpha$/H$\beta$ and H$\gamma$/H$\beta$ samples, respectively.

We have shown that there are statistically significant anomalous Balmer line ratios in our sample, indicating a population of galaxies which deviate from Case B. However, it is impossible to distinguish between the scenarios which can cause deviations from Case B (e.g. density-bounded nebulae or optically thick Balmer lines) by considering individual Balmer line ratios, as each model can produce a spread of these values. Instead, constraining power is strongest for galaxies with multiple Balmer line detections. For this reason, in this paper we focus our attention on the subsample of galaxies with a $5\sigma$ detection of H$\alpha$, H$\beta$, and H$\gamma$. While this leaves us with a smaller sample, the distribution of galaxies on the H$\alpha$/H$\beta$ versus H$\gamma$/H$\beta$ plot shown in Fig.~\ref{fig:Ha_Hg_vs_Ha_Hb_caseB} will allow us to disentangle dust and different non-Case B scenarios.

To demonstrate this, we have also plotted lines representing Balmer absorption due to a shell of excited hydrogen surrounding a Case B ionisation-bounded nebula, following the prescription of \citet{Yanagisawa:2024ac}. We can see that while the combination of gas which is optically thick to Balmer lines and dust attenuation can explain a number of ABEs, this scenario cannot account for them all. In particular, galaxies which have elevated H$\gamma$/H$\beta$ are not well described by this scenario.

\subsection{Density-bounded nebulae models}
\label{sec:Density-bounded nebulae models} 

We employ the spectral synthesis code \texttt{CLOUDY} \citep{Ferland:1998aa, Chatzikos:2023aa} to simulate the emission from both density-bounded and ionisation-bounded nebulae. We adopt a spherical gas model with an inner radius of 10\,pc and a constant gas density. The calculation stops when an electron fraction of 0.01 is reached, or if the temperature falls below 3000\,K. For density-bounded models, we apply an additional stopping condition based on the cloud depth. The SEDs for the ionising source in our modelling are an instantaneous starburst simple stellar population (SSP) from BPASS v2.2.1 with \citet{Chabrier:2003aa} initial mass function (IMF) with a $300\,\Msun$ cutoff and including binaries \citep{Eldridge:2017aa, Stanway:2018aa}. 

\begin{table}
	\centering
	\caption{The parameter values for the \texttt{CLOUDY} grid of models. The models have spherical geometry and a constant gas density. The stellar SEDs are from BPASS v2.2.1 and include binaries \citep{Eldridge:2017aa, Stanway:2018aa}. The luminosity of the SSP is scaled via the ionisation parameter. The abundances are scaled with metallicity based on solar abundances \citep{Asplund:2009aa}.}
	\label{tab:cloudy_parameters}
	\begin{tabular}{cc} % four columns, alignment for each
		\hline
		Parameter & Values \\
		\hline
        Gas density, $\mathrm{n_H}$ [$\textrm{cm}^{-3}$] & $10^{1}$, $10^{2}$, $10^{3}$, $10^{4}$ \\
        Metallicity,  $\log\,Z/\Zsun$ & $-3$, $-2.5$, $-2$, $-1.8$, $-1.6$, $-1.4$, $-1.2$,\\
        & $-1$, $-0.8$, $-0.6$, $-0.4$, $-0.2$, $0$, $0.2$\\
        Ionisation parameter, $\log\,\mathrm{U}$ & $-3$, $-2.5$, $-2$, $-1.5$, $-1$ \\
        Stellar ages [Myr] & $1$, $3$, $5$, $10$ \\
		\hline
	\end{tabular}
\end{table}

The free parameters for our models are gas density ($\mathrm{n_H}$), metallicity ($\log Z/\Zsun$), ionisation parameter ($\log\,\mathrm{U}$), and stellar age. We consider a grid of parameter values, which is given in Table~\ref{tab:cloudy_parameters}. The gas-phase and stellar metallicity are always equivalent unless explicitly stated otherwise. The chemical abundances are scaled with the metallicity based on solar abundances \citep{Asplund:2009aa}.

\subsection{Comparison to observed line ratios}
\label{sec:Comparison to observed line ratios} 

\begin{figure}
    \centering
	\includegraphics[width=\columnwidth]{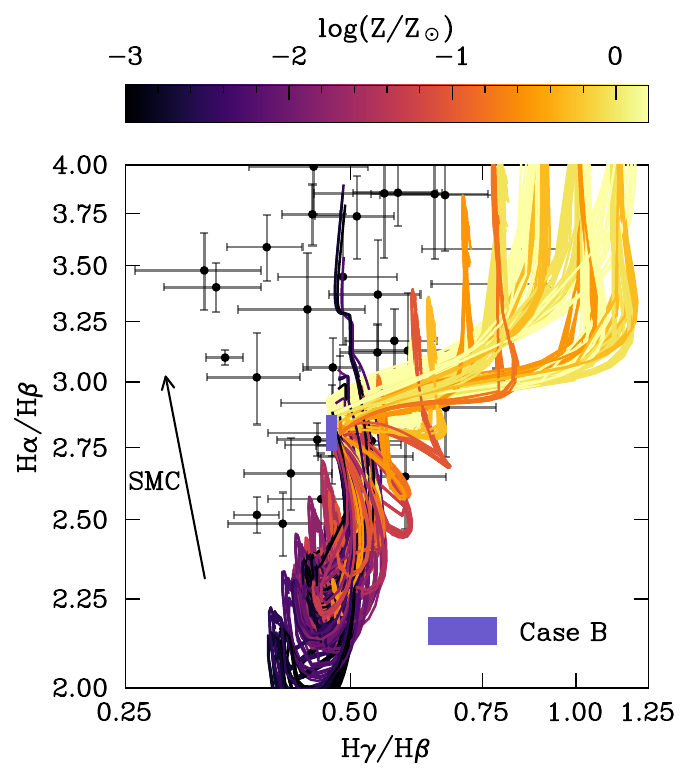}
    \caption{H$\alpha$/H$\beta$ against H$\gamma$/H$\beta$ for all density-bounded nebulae models in our parameter grid (Table~\ref{tab:cloudy_parameters}). Each line represents a density-bounded model with increasing cloud depth until it becomes ionisation-bounded and converges to the Case B line ratios (blue box shaded region around H$\alpha$/H$\beta=2.8$ and H$\gamma$/H$\beta=0.47$).  Our models bracket the bottom right of the observed distribution, and therefore can account for all observed values when combined with dust attenuation (arrow shows \citealp{Gordon:2003aa} SMC extinction curve with a length of $\mathrm{A}_{\rm _V}=1$ mag). The model-to-model variation is primarily driven by metallicity, and the trends are otherwise robust to changes in gas density, ionisation parameter, and stellar age, although the older 10\,Myr stars tend to reach more extreme line ratios while still following the same trends set by the metallicity. }
    \label{fig:DB_Ha_Hg_vs_Ha_Hb_allmod}
\end{figure}

In Fig.~\ref{fig:DB_Ha_Hg_vs_Ha_Hb_allmod} we show all models produced from our grid of parameters (Table~\ref{tab:cloudy_parameters}) and we compare them to the observed sample. The trends of our density-bounded nebulae models are remarkably robust to changes in gas density, ionisation parameter, and stellar age. Metallicity is primary parameter which causes changes in the H$\alpha$/H$\beta$ versus H$\gamma$/H$\beta$ trends. We note that changing the stellar age in the range we consider can act to shift the metallicity trend by a factor of $\sim2$, however this is a minor effect overall compared to the impact of changing metallicity (see Appendix~\ref{sec:Impact of stellar age}). At the oldest stellar ages we consider, 10\,Myr, the models can also reach noticeably more extreme values for the line ratios reaching values of H$\alpha$/H$\beta<2$ at low metallicities.

While metallicity is the dominant driver of variation between models, the effect of varying metallicity between $Z=0.001\,$--$\,0.1\,\Zsun$ is relatively minor, mainly extending the models to even lower values of H$\alpha$/H$\beta$. Models in this metallicity range are generally increasing in both H$\alpha$/H$\beta$ and H$\gamma$/H$\beta$ with increasing cloud depth. Varying metallicity in the range of $Z=0.1\,$--$\,1\,\Zsun$ leads to a stark turnover in the H$\alpha$/H$\beta$ against H$\gamma$/H$\beta$ trend. Models with intermediate metallicities, $\log\,(Z\,[\Zsun])\approx-0.5$, start with decreasing H$\alpha$/H$\beta$ and increasing H$\gamma$/H$\beta$ with increasing cloud depth, before reversing on themselves and converging on the Case B values. At solar metallicity, models decrease in both H$\alpha$/H$\beta$ and H$\gamma$/H$\beta$ with increasing cloud depth until converging on the Case B values. This turnover means that density-bounded models can naturally explain the variety of scatter around the Case B values and the lack of an obvious trend. Our density-bounded models also explain why galaxies with the lowest H$\alpha$/H$\beta$ ratios also have low H$\alpha$/H$\gamma$. 

The metallicity variation between models in Fig.~\ref{fig:DB_Ha_Hg_vs_Ha_Hb_allmod} is driven almost entirely by stellar metallicity, rather than by gas-phase metallicity. We tested this by rerunning our models twice, keeping either gas-phase or stellar metallicity fixed. We found that there was a small dependence on the gas-phase metallicity, which we believe is due to the changing temperature. However, the observed turnover in Fig.~\ref{fig:DB_Ha_Hg_vs_Ha_Hb_allmod} is present even when gas-phase metallicity is fixed. We further discuss the variation of the Balmer line ratios with cloud depth, and why stellar metallicity is important in  Section~\ref{sec:Anatomy of density-bounded nebulae}.

\subsection{Anatomy of density-bounded nebulae}
\label{sec:Anatomy of density-bounded nebulae} 

\begin{figure*} 
\centering
	\includegraphics[width=\textwidth]{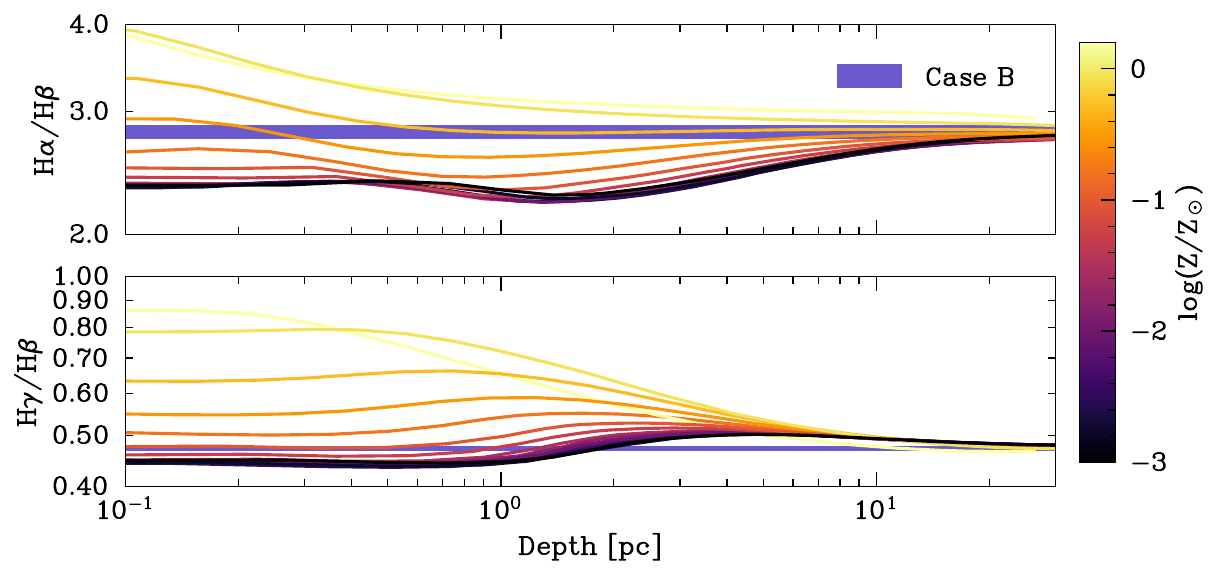}
    \caption{The H$\alpha$/H$\beta$ and H$\gamma$/H$\beta$ ratios as a function of cloud depth for a set of density-bounded nebulae models. These represent gas with density $\mathrm{n_H}=10~\mathrm{cm^{-3}}$ being ionised by a 3 Myr binary instant starburst with $\log(\mathrm{U})=-1$. The $\log\mathrm{Z/\Zsun}$ is varied from $-3$ to 0.2. All models show a varying H$\alpha$/H$\beta$ and H$\gamma$/H$\beta$ with increasing cloud depth, which is caused by the transition from the gas being optically thin to optically thick for Lyman lines. For low metallicity models, both H$\alpha$/H$\beta$ and H$\gamma$/H$\beta$ initially have low values, before converging to Case B as they become ionisation-bounded nebulae. However, high metallicity models begin with high H$\alpha$/H$\beta$ and H$\gamma$/H$\beta$, and converge down to the Case B values as they become ionisation bounded. This is driven by the change Lyman line absorption features with stellar metallicity, in particular the increasing absorption features around Ly$\gamma$ with increasing metallicity (Fig.~\ref{fig:bpass_plot}).}
    \label{fig:DB_radial_plot}
\end{figure*}

In Fig.~\ref{fig:DB_radial_plot} we show the variation of H$\alpha$/H$\beta$ and H$\gamma$/H$\beta$ as a function of cloud depth. As cloud depth increases, there is a transition from being optically thin to Lyman lines (Case A and Case C) to optically thick to Lyman lines (Case B and Case D).

In Section~\ref{sec:Comparison to observed line ratios}, we showed that there is a turnover in the H$\alpha$/H$\beta$ versus H$\gamma$/H$\beta$ relation with varying stellar metallicity. This is driven by the Lyman absorption features in the stellar SEDs. In Fig.~\ref{fig:bpass_plot}, we show a comparison of the SEDs of 3\,Myr SSP with metallicities of $Z=0.001$ and $Z=\Zsun=0.014$. Higher metallicity stars have much larger absorption around Ly$\gamma$ than their lower metallicity counterparts. This is vitally important to the H$\beta$ flux, because Ly$\gamma$ continuum pumping causes a transition from $\mathrm{n}=1$ to $\mathrm{n}=4$. The electron in the $\mathrm{n}=4$ state can then undergo spontaneous radiative decay to $\mathrm{n}=2$, emitting a H$\beta$ photon. For high-metallicity stars, Ly$\beta$ continuum pumping is stronger relative to Ly$\gamma$ pumping, so the H$\alpha$/H$\beta$ values are above Case B.

Therefore, given that continuum pumping is important in our models, we can say that they transition from Case C to Case D with increasing cloud depth. When our models become ionisation bounded they can be reasonably approximated as Case B. This is because the continuum pumping is a small effect when the gas thickness is large due to the fact that Lyman line photons are used up at small depths relative to the total depth of an ionisation-bounded cloud \citep{Luridiana:2009aa}. This is shown as our models all converge on the Case B values in Fig.~\ref{fig:DB_Ha_Hg_vs_Ha_Hb_allmod}.

Of course, the physical scale of our density-bounded models is sensitive to the ionisation parameter and density. As previously noted, changing these parameters does not have a large effect on the H$\alpha$/H$\beta$ versus H$\gamma$/H$\beta$ trends, and therefore the line ratio trends scale with the total cloud depth. Although the general trends of high-density gas and a lower luminosity source leading to smaller density-bounded nebulae is expected, the actual physical sizes of our models should not be taken literally. These models could be viewed as individual density-bounded clouds, the physical scale of which can be changed drastically or represent an ensemble of clouds in a realistic galactic environment.

We also test the effect of including microturbulent velocity, $v_{\rm turb}$, which broadens the Doppler width of a line in conjunction with the thermal motion $b=\sqrt{v_{\rm th}^2 + v_{\rm turb}^2}=\sqrt{2}\sigma$. Turbulence can affect the pumping of transitions since a broader continuum range is able to pump a transition. Locally, \ion{H}{II} regions show velocity dispersion in the range of 18--30 $\mathrm{km\,s^{-1}}$, and diffuse gas shows velocity dispersion in the range of 20--60 $\mathrm{km\,s^{-1}}$ \citep{Law:2022aa}. Additionally, these velocity dispersions tend to increase with star formation rate surface density, and we see a significant amount of recent star formation in ABEs (see Section~\ref{sec:Continuum features and Lyman-alpha}), and therefore we may expect the velocity dispersion of ABEs to be on the high end. We must also of course consider that velocity dispersion increases significantly with redshift \citep{Ubler:2019aa}. Therefore, a velocity dispersion of 50--100 $\mathrm{km\,s^{-1}}$ may be more appropriate for galaxies at $z>2$ \citep{Ubler:2023aa, Arribas:2024aa, de-Graaff:2024aa, Carniani:2024aa}. However, if these density-bounded nebulae are in fact entire-galaxy systems, rather than a few of the brightest small-scale nebulae (we discuss this possibility in Section~\ref{sec:Physical origin of density-bounded nebulae}), then the larger scale bulk motion of the gas may become important as well as the velocity dispersion. It is important to note that these bulk motions likely have correlated velocity structure and act as non-thermal turbulence \citep{Padoan:2012aa}, and that turbulence can effect the way ionising photons escape a nebula \citep{Kakiichi:2021aa}. The full impact of these effects is beyond the scope of our simplified modelling, and so we use microturbulence as a proxy. The width $\sigma_{\rm obs}$ (i.e. the combination of velocity dispersion and beam-smeared bulk motion) of narrow Balmer lines in galaxies at $z\approx6$ can be > 150 $\mathrm{km\,s^{-1}}$, with outflows having significantly higher values > 500 $\mathrm{km\,s^{-1}}$ \citep{Ubler:2023aa, Nelson:2024aa, Arribas:2024aa, Jones:2024aa}. 

Due to this large range in reasonable values, we test a wide range of $\sigma_{\rm turb}$ turbulence values: 14, 70, and 140 $\mathrm{km\,s^{-1}}$. However, while we found that the physical scale of the model could be increased by a factor of $\sim$2, the Balmer line ratios were unchanged. If the idealised large-scale microturbulence proxy holds, then this scenario would be valid on a much larger physical scale, such as an entire galaxy, where the velocity dispersion can be significant.

We show the effect of turbulence in Fig.~\ref{fig:D_FESC}, where we use the LyC $f_{\mathrm{esc}}$. We use $f_{\mathrm{esc}}$ rather than depth due to the problems in comparing depths between models discussed earlier. Models with very little turbulence (14 $\mathrm{km\,s^{-1}}$) become consistent with Case B when the ionising escape is very high, $f_{\mathrm{esc}}>0.8$. However, increasing the turbulence to a more reasonable value, 70 $\mathrm{km\,s^{-1}}$, allows the H$\alpha$/H$\beta$ ratio to remain below Case B for significantly lower values of $f_{\mathrm{esc}}$ due to the increased range of continuum which can pump the Lyman lines. Stellar age can also be important for the H$\alpha$/H$\beta$ versus $f_{\mathrm{esc}}$ relationship because it changes the strength of the ionising continuum relative to the continuum around the Lyman lines.

We also note that the presence of turbulence can alleviate concerns about density-bounded models producing too-small equivalent widths, such a those outlined in \citet{Scarlata:2024aa}. We show this effect in Fig.~\ref{fig:D_FESC}. With increased turbulence, the EW[H$\beta$] is increased slightly at a given $f_{\mathrm{esc}}$, but more importantly the H$\alpha$/H$\beta$ remains sub-Case B deeper into the cloud (i.e. for lower values of $f_{\mathrm{esc}}$), and therefore anomalous line ratios would be observed at larger equivalent widths. For a more specific example, if we include $70\,\mathrm{km\,s^{-1}}$ turbulence for a model with $\mathrm{n_H}=10~\mathrm{cm^{-3}}$, $\log(\mathrm{U})=-2.5$, and $Z = 0.1\,\Zsun$, then the equivalent width of H$\beta$ is EW[H$\beta]=181$\AA\ when the H$\alpha$/H$\beta$ value is 2.63. This model would therefore be consistent with the values seen in SXDF308, which has H$\alpha$/H$\beta=2.62\pm0.08$ and EW[H$\beta]=165\pm14$\AA\ \citep{Scarlata:2024aa}. However, even without turbulence, it is possible to get reasonable equivalent widths for many parameters.  In any case, we are hesitant to rule out any of the parameter space using equivalent widths from \texttt{CLOUDY} models alone due to the fact that the geometry of galaxies showing density-bounded emission is likely complex, and we discuss this further in Section~\ref{sec:Alternative explanations for anomalous Balmer line ratios}.

\section{Other spectral features}
\label{sec:Other spectral features} 

In this section we discuss other spectral features observed in ABEs and how they relate to density-bounded nebulae. In Section~\ref{sec:Continuum features and Lyman-alpha} we analyse the Balmer continuum, UV slopes, and Ly$\alpha$ emission of the observed galaxies and our density-bounded models. In Section~\ref{sec:Metal line ratios} we consider the [\ion{O}{III}]/[\ion{O}{II}] and [\ion{O}{III}]/H$\beta$ ratios.

\subsection{Continuum features and Lyman-alpha}
\label{sec:Continuum features and Lyman-alpha} 

\begin{figure} 
\centering
	\includegraphics[width=\columnwidth]{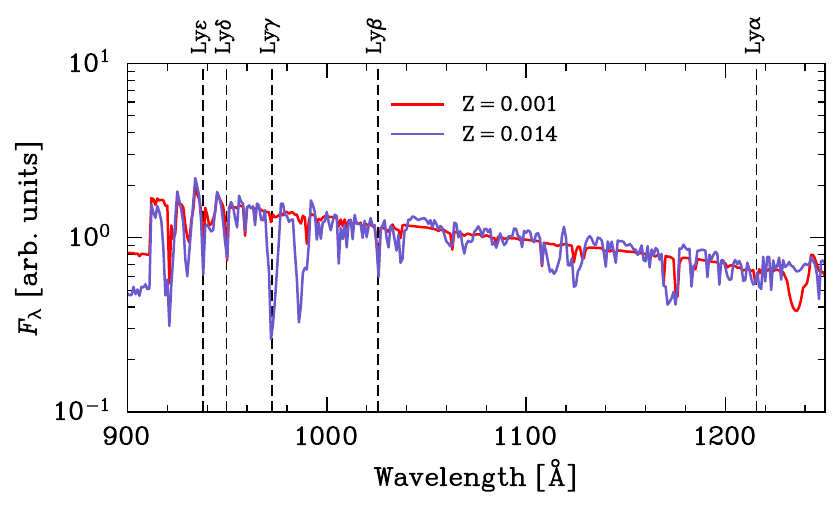}
    \caption{A comparison of the SEDs of 3\,Myr SSP with metallicities of $\mathrm{Z}=0.001$ (red) and $\mathrm{Z}=0.014$ (blue) from the BPASS library. The Lyman line absorption features vary with metallicity. Notably, the higher metallicity stars have much stronger absorption around Ly$\gamma$, and therefore fewer continuum photons are available to pump electrons to the $\mathrm{n}=4$ state of hydrogen and cause H$\beta$ emission via subsequent radiative decay. The variation in absorption features is responsible for the turnover in the H$\alpha$/H$\beta$ versus H$\gamma$/H$\beta$ relation shown in Fig.~\ref{fig:DB_Ha_Hg_vs_Ha_Hb_allmod}. These absorption features are not important in ionisation-bounded nebulae because the continuum photons which can pump Lyman lines are used up at smaller cloud depths than the ionising photons, meaning that most emission is from non-pumped gas.}
    \label{fig:bpass_plot}
\end{figure}

\begin{figure} 
\centering
	\includegraphics[width=\columnwidth]{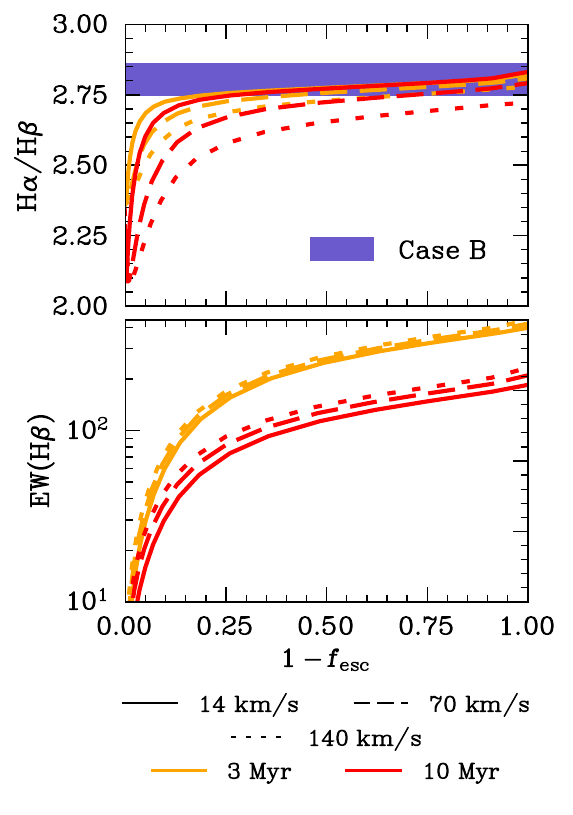}
    \caption{The H$\alpha$/H$\beta$ ratio and EW[H$\beta$] as a function of $1-f_{\mathrm{esc}}$ for a set of density-bounded nebulae models. These models represent gas with density $\mathrm{n_H}=10~\mathrm{cm^{-3}}$ being ionised by a 3 or 10 Myr binary instant starburst with $\log(\mathrm{U})=-1$. The $\log\mathrm{Z/\Zsun}$ is fixed at $-2$. The models include microturbulent velocity dispersions $\sigma_{turb}$ of 14 (solid), 70 (dashed), or 140 (dotted) $\mathrm{km\,s^{-1}}$. Although the H$\alpha$/H$\beta$ versus H$\gamma$/H$\beta$ trends in Fig.~\ref{fig:DB_Ha_Hg_vs_Ha_Hb_allmod} are fairly robust to changes in stellar age, either shifting the metallicity values or making the trends more extreme, stellar age can have a strong effect on the evolution of H$\alpha$/H$\beta$ with $f_{\mathrm{esc}}$ and on the scaling of EW[H$\beta$]. With very little turbulence, the models become consistent with Case B while $f_{\mathrm{esc}}>0.8$. However, adding significant but reasonable turbulence pushes this to $f_{\mathrm{esc}}=0.6$ and $f_{\mathrm{esc}}=0.3$ for the 3\,Myr and 10\,Myr stars respectively. In the most extreme case, with 10\,Myr stars and $\sigma_{turb}=140\,\mathrm{km\,s^{-1}}$, the model shows sub-Case B H$\alpha$/H$\beta$ even when it is ionisation bounded.}
    \label{fig:D_FESC}
\end{figure}

\begin{figure} 
\centering
	\includegraphics[width=\columnwidth]{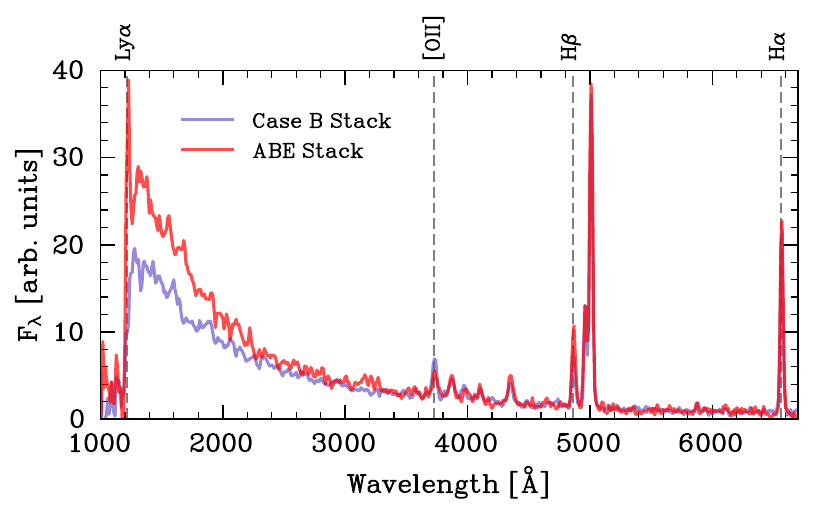}
    \caption{Median stacked spectra of ABEs (red) and galaxies consistent with Case B H$\alpha$/H$\beta$ and no dust attenuation (blue). The ABEs were selected as those with H$\alpha$/H$\beta$ at least 1$\sigma$ below Case B, with 14 galaxies in this sample. The Case B sample was chosen as those at least 1$\sigma$ above the H$\alpha$/H$\beta$ Case B lower limit, but not more that 1$\sigma$ above the H$\alpha$/H$\beta$ Case B upper limit, with 9 galaxies in this sample. These selection criteria were used to avoid any galaxies which were ambiguous and to ensure that the Case B sample did not include galaxies with significant dust attenuation. The ABE stack has features consistent with density-bounded nebulae, such as the presence of Ly$\alpha$ and little [\ion{O}{II}] emission. Notably, the ABE stack also shows a steep UV slope ($\beta=-2.40\pm0.08$ compared to $\beta=-2.02\pm0.07$ for the Case B stack) and a Balmer jump ($F_{\nu,4225}/F_{\nu,3565}=0.78\pm0.08$ compared to $F_{\nu,4225}/F_{\nu,3565}=1.02\pm0.06$ for the Case B stack). The low [\ion{O}{II}] emission, steep UV slope, and Ly$\alpha$ emission seen in the ABE stack are consistent with density-bounded nebulae.}
    \label{fig:JADES_Ha_Hb_vs_BJ}
\end{figure}

In Fig.~\ref{fig:JADES_Ha_Hb_vs_BJ} we show the median stacked spectra for our sample of $z>5.3$ galaxies with low H$\alpha$/H$\beta$, ABEs, compared to a sample of Case-B consistent galaxies. The ABE sample is 14 galaxies with at least 1$\sigma$ below the Case B lower limit for H$\alpha$/H$\beta$. The Case B sample is 9 galaxies with at least 1$\sigma$ above the H$\alpha$/H$\beta$ Case B lower limit, but not more that 1$\sigma$ above the H$\alpha$/H$\beta$ Case B upper limit. This means that both samples exclude galaxies which are ambiguous, i.e. galaxies with H$\alpha$/H$\beta$ ratios which may or may not be consistent with Case B. Due to the tight range selected for the Case B galaxies, dust attenuation should not be a significant factor. There is of course the risk that dust extinction could move some density-bounded galaxies into the Case B stack, or that some higher metallicity density-bounded galaxies would fall into the Case B range, so the comparison is not completely clean of contamination. However, we can be sure that no Case B consistent galaxies can enter the ABE stack. 

Median stacking has proven effective in numerous previous studies at extracting the spectral properties of a population of galaxies \citep[e.g.][]{Harikane:2020aa, Witten:2023aa, Roberts-Borsani:2024aa}. In order to create a stack of our ABE and Case-B consistent galaxies, we follow a similar process to \citet{Witten:2025aa}. We first shift the spectra of each constituent galaxy to rest-frame and rebin to a common wavelength grid, with a resolution of 0.001 $\mu$m, using the SpectRes code \citep{Carnall:2017aa}. Each spectrum is normalised such that the flux at $0.15 \mu \mathrm{m}$ is unity, and we then perform a median stacking. We estimate the uncertainties of our stack with jackknife sampling by generating 1000 stacks by randomly removing one of the galaxies from our sample. Each time we select a galaxy for stacking we redraw the galaxy's spectrum from Gaussian's centred on the observed flux in a given wavelength bin and the standard deviation of the Gaussian is given by the uncertainty in the flux. We then determine the median of these stacked spectra and take the standard deviation of these stacks as our reported uncertainties in the spectrum. The median redshift of the ABE stack is $z=6.44$, compared to $z=5.89$ for the Case B sample. The lower limit of the redshift is chosen as $z>5.3$ and the upper limit is $z<\sim7$ based on the observability of H$\alpha$. We measured $M_{\mathrm{UV}}$ based on the median flux between 1400 and 1600 \AA\ in the spectra and the spectroscopic redshift. We find a median and $16^\text{th}$--$84^\text{th}$ percentile scatters of $M_{\mathrm{UV}}=-19.6^{+0.7}_{-0.3}$ mag for the Case B sample and $M_{\mathrm{UV}}=-19.0^{+0.9}_{-0.5}$ mag for the anomalous sample.

We fit the UV slope with a power law $\lambda^\beta$ in the wavelength range $\lambda=$\,[1200--2600]\,\AA using the wavelength windows defined by \citet{Calzetti:1994aa}.
The UV slope of the galaxies with sub-Case B H$\alpha$/H$\beta$ is $\beta=-2.40\pm0.08$ compared to $\beta=-2.02\pm0.07$ for the Case B stack. The Case B stack has a UV slope more typical of bright galaxies ($M_{\mathrm{UV}}\approx-20.2$) at $z=6$, whereas the ABE stack is consistent with faint galaxies ($M_{\mathrm{UV}}\approx-17.8$) at this redshift by \citet{Topping:2024ab}. We should note that due to our calculation of $M_{\mathrm{UV}}$ from spectra rather than photometry, our measured $M_{\mathrm{UV}}$ is not directly comparable to \citet{Topping:2024ab} as there many be an offset due to slit losses. 

Following \citet{Roberts-Borsani:2024aa}, we calculate the Balmer jump/break strength of these galaxies using two wavelength windows of $\lambda_{\mathrm{rest}}=3500$--$3630$ \AA\ and 4160--4290 \AA\ to avoid contamination by emission lines. The ABE stack shows a Balmer jump, with $F_{\nu,4225}/F_{\nu,3565}=0.78\pm0.08$, whereas the Case B sample shows $F_{\nu,4225}/F_{\nu,3565}=1.02\pm0.06$. The Case B sample does not indicate either a Balmer jump or Balmer break strongly, and is typical for galaxies observed at $z\sim6$. However, the Balmer jump shown in the ABE stack is clearly significant; Balmer jumps of this magnitude are not typical even at $z\sim9$ \citep{Roberts-Borsani:2024aa}.

Steeper UV slopes tend to indicate younger stellar populations, which is certainly consistent with the observed Balmer jump. However, \citet{Topping:2024ab} show that galaxies with extreme UV slopes ($\beta<-2.8$), photoionisation models which allow for Lyman continuum (LyC) leakage (density-bounded nebulae) are preferred. While our ABE stack does not have a UV slope which is nearly as extreme as the extreme \citet{Topping:2024ab} galaxies, the ABE stack does have a significantly steeper slope than the Case B stack, which is consistent with density-bound emission.

The ABE stack also shows clear Ly$\alpha$ emission. This is consistent with a density-bounded scenario. Ionisation-bounded nebulae are surrounded by neutral material which scatters Ly$\alpha$, increasing the likelihood that it will be destroyed by dust or converted to two-photon continuum unless there are low-density channels available for the Ly$\alpha$ to escape. On the other hand, density-bounded nebulae more naturally allow for Lyman line escape as the ionised nebulae cannot be (immediately) surrounded by significant columns of neutral gas. Additionally, the intrinsic Ly$\alpha$/H$\beta$ of density-bounded models can be significantly changed relative to the Case B, with the ratio being increased when LyC $f_{\mathrm{esc}}$ is high ($\sim0.9$) and lower for $f_{\mathrm{esc}}<0.9$ until the nebulae becomes ionisation bounded. However, this does not mean that galaxies which show density-bounded emission \textit{must} show Ly$\alpha$ emission; the geometry of galaxies is complicated and while the emission from a galaxy may be dominated by density-bounded nebulae, the Ly$\alpha$ may still be scattered or destroyed in the circumgalactic medium (CGM), the intergalactic medium (IGM), or even clumps in the ISM. We also note that LAEs observed at high redshifts ($z>7$, where H$\alpha$ is no longer directly measurable with \textit{JWST} NIRSpec) have properties consistent with ABEs \citep{Witstok:2024ab}.

\subsection{Metal line ratios}
\label{sec:Metal line ratios} 

\begin{figure*} 
\centering
	\includegraphics[width=\textwidth]{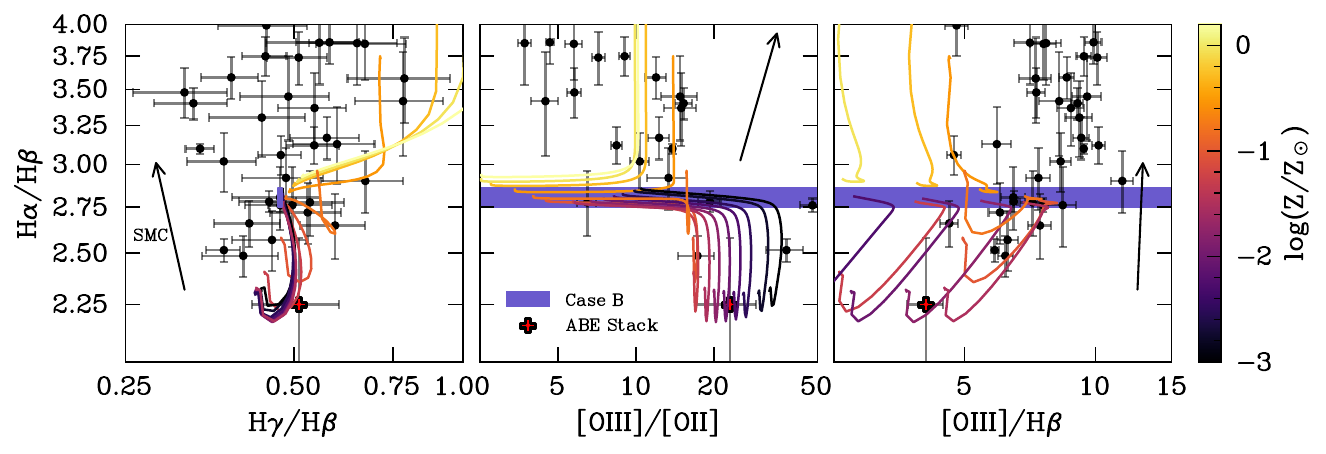}
    \caption{H$\alpha$/H$\beta$ against H$\gamma$/H$\beta$, [\ion{O}{III}]/[\ion{O}{II}], and [\ion{O}{III}]/H$\beta$ for a select set of density-bounded nebulae models. These models are spherical \texttt{CLOUDY} photoionisation models of gas with $\mathrm{n_H}=10^3\mathrm{cm^{-3}}$ being ionised by with a 5 Myr binary instant starburst with $\log(\mathrm{U})=-2.5$. The $\log\mathrm{Z/\Zsun}$ is varied from $-3$ to 0.2. JADES galaxies at $z>5.3$ with 5$\sigma$ detections of H$\alpha$, H$\beta$, and H$\gamma$ are plotted in black, and the ABE stack line ratios are shown with a large red cross. The arrow represents increasing dust attenuation for an SMC extinction curve ($\mathrm{R}_{\rm _V}=2.7$) with a length of $\mathrm{A}_{\rm _V}=1$ mag \citep{Gordon:2003aa}. These parameters were selected to show that density-bounded nebulae and dust attenuation are able to reproduce the observations for all galaxies with a sub-Case B H$\alpha$/H$\beta$ without needing to resort to extreme parameters. Slight changes in the ionisation parameter can increase or decrease [\ion{O}{III}]/[\ion{O}{II}]. Our models can reproduce the [\ion{O}{III}]/H$\beta$ ratios of sub-Case B galaxies well. Accounting for dust attenuation, we can also reproduce most galaxies with higher H$\alpha$/H$\beta$.}
    \label{fig:DB_cloudy_obs_comp}
\end{figure*}

Anomalous Balmer line ratios are far from the only interesting spectral feature observed in high-redshift galaxies and their low-redshift analogues. Metal line ratios contain a wealth of information about the chemical evolution and ionisation structure of galaxies. If density-bounded nebulae are indeed prevalent in the early Universe, their fingerprints may be visible in these ratios.

The [\ion{O}{III}]/[\ion{O}{II}] ratio is particularly interesting due to its evolution with redshift. Galaxies in the early Universe have higher values of [\ion{O}{III}]/[\ion{O}{II}] than those observed locally, the cause of which has been interpreted as due to high ionisation parameter, high electron density, or hard ionising spectra \citep{Cameron:2023aa,Scholtz:2023aa,Reddy:2023aa,Topping:2024aa}.

Density-bounded nebulae naturally lead to higher [\ion{O}{III}]/[\ion{O}{II}] ratios for a given ionisation parameter, electron density, and ionising SED \citep{Zackrisson:2013aa}. This is because the inner parts of a nebula have a higher $\mathrm{O^{++}}$/$\mathrm{O^{+}}$ ratio than the outer regions, and therefore we expect most [\ion{O}{III}] emission to occur in the central parts of the nebula. Density-bounded models terminate before significant [\ion{O}{II}] emission has occurred. This effect can be seen in Fig.~\ref{fig:DB_cloudy_obs_comp}, where we show H$\alpha$/H$\beta$ against H$\gamma$/H$\beta$, [\ion{O}{III}]/[\ion{O}{II}], and [\ion{O}{III}]/H$\beta$ for an ordinary set of parameter values: $\mathrm{n_H}=10^3\mathrm{cm^{-3}}$, $\log(\mathrm{U})=-2.5$, and a 5 Myr ionising SED. To account for the effects of $\alpha$-enhancement on oxygen line emission in low-metallicity galaxies, we have enhanced the oxygen abundance based on the iron abundance, following \citet{Nicholls:2017aa}. For small cloud depths, the [\ion{O}{III}]/[\ion{O}{II}] ratio is higher. The [\ion{O}{III}]/H$\beta$ trends are less obvious due to the excess H$\beta$ emission in our density-bounded models.

In Fig.~\ref{fig:DB_cloudy_obs_comp} we also show H$\alpha$/H$\beta$ against H$\gamma$/H$\beta$, [\ion{O}{III}]/[\ion{O}{II}], and [\ion{O}{III}]/H$\beta$ for JADES galaxies at $z>5.3$ with 5$\sigma$ detections of H$\alpha$, H$\beta$, and H$\gamma$. We can see a tentative trend of increasing [\ion{O}{III}]/[\ion{O}{II}] with decreasing H$\alpha$/H$\beta$. This is expected if anomalous Balmer line ratios are driven by density-bounded models, and is the opposite trend to what is expected for Case B and dust attenuation. This is further confirmed by the comparison of our anomalous and Case B stacks in Fig.~\ref{fig:JADES_Ha_Hb_vs_BJ}, where we can see that the galaxies with sub-Case B H$\alpha$/H$\beta$ show significantly less [\ion{O}{II}] emission.

\citet{Scarlata:2024aa} consider local galaxies in the SDSS survey, and associate high [\ion{O}{III}]/[\ion{O}{II}] emission with anomalous Balmer line ratios. In particular they find that high [\ion{O}{III}]/[\ion{O}{II}] galaxies have high H$\gamma$/H$\beta$ and low H$\alpha$/H$\beta$. This is consistent with the density-bounded models. The low H$\alpha$/H$\beta$ galaxies will be dominated by density-bounded nebulae, and thus have higher [\ion{O}{III}]/[\ion{O}{II}] on average. While low metallicity density-bounded nebulae can have sub-Case B H$\gamma$/H$\beta$, overall the sub-Case B H$\gamma$/H$\beta$ population is likely dominated by ordinary galaxies with dust attenuation. Instead, we would expect high H$\gamma$/H$\beta$ values to be dominated by density-bounded nebulae, and thus to have high [\ion{O}{III}]/[\ion{O}{II}], as is observed.

Our models can reproduce the [\ion{O}{III}]/H$\beta$ ratios of sub-Case B galaxies reasonably well. Accounting for dust attenuation, we can also reproduce nearly all galaxies with higher H$\alpha$/H$\beta$. We note that agreement with the very highest [\ion{O}{III}]/H$\beta$ ratios is difficult for all one-component models, and seems more pronounced for ionisation-bounded models than for the density-bounded models shown here \citep{Cameron:2023aa, Roberts-Borsani:2024aa}.

For similar reasons as for [\ion{O}{III}]/[\ion{O}{II}], our models create high [\ion{O}{III}]$88\,\mu$m / [\ion{C}{II}]$158\,\mu$m ratios. We do not have sufficient data to quantitatively compare our models, but we note that this is qualitatively similar to the high ratios seen in high-redshift galaxies \citep{Hashimoto:2019aa,Harikane:2020ab,Carniani:2020aa,Witstok:2022aa}. Additionally, [\ion{O}{III}]$88\,\mu$m has a low critical density, and is therefore enhanced in our models with low density. Again, we do not have sufficient data to compare our models, but this is a potential indication that the density-bounded emission arises from low-density gas rather than high-density clumps.

\section{Discussion}
\label{sec:Discussion}

In this section we discuss the implications of density-bounded nebulae on our understanding of galaxies in the early Universe. In Section~\ref{sec:Physical origin of density-bounded nebulae} we consider how density-bounded emission may arise in galaxies due to burst-quench cycles in high-redshift galaxies and local dwarfs. In Section~\ref{sec:Complications arising from density-bounded nebulae} we explore the complications which may need to be considered if a significant number of high-redshift galaxies are density-bounded. In Section~\ref{sec:Alternative explanations for anomalous Balmer line ratios} we discuss alternative explanations for ABEs.

\subsection{Physical origin of density-bounded nebulae}
\label{sec:Physical origin of density-bounded nebulae} 

\begin{figure*} 
\centering
	\includegraphics[width=\textwidth]{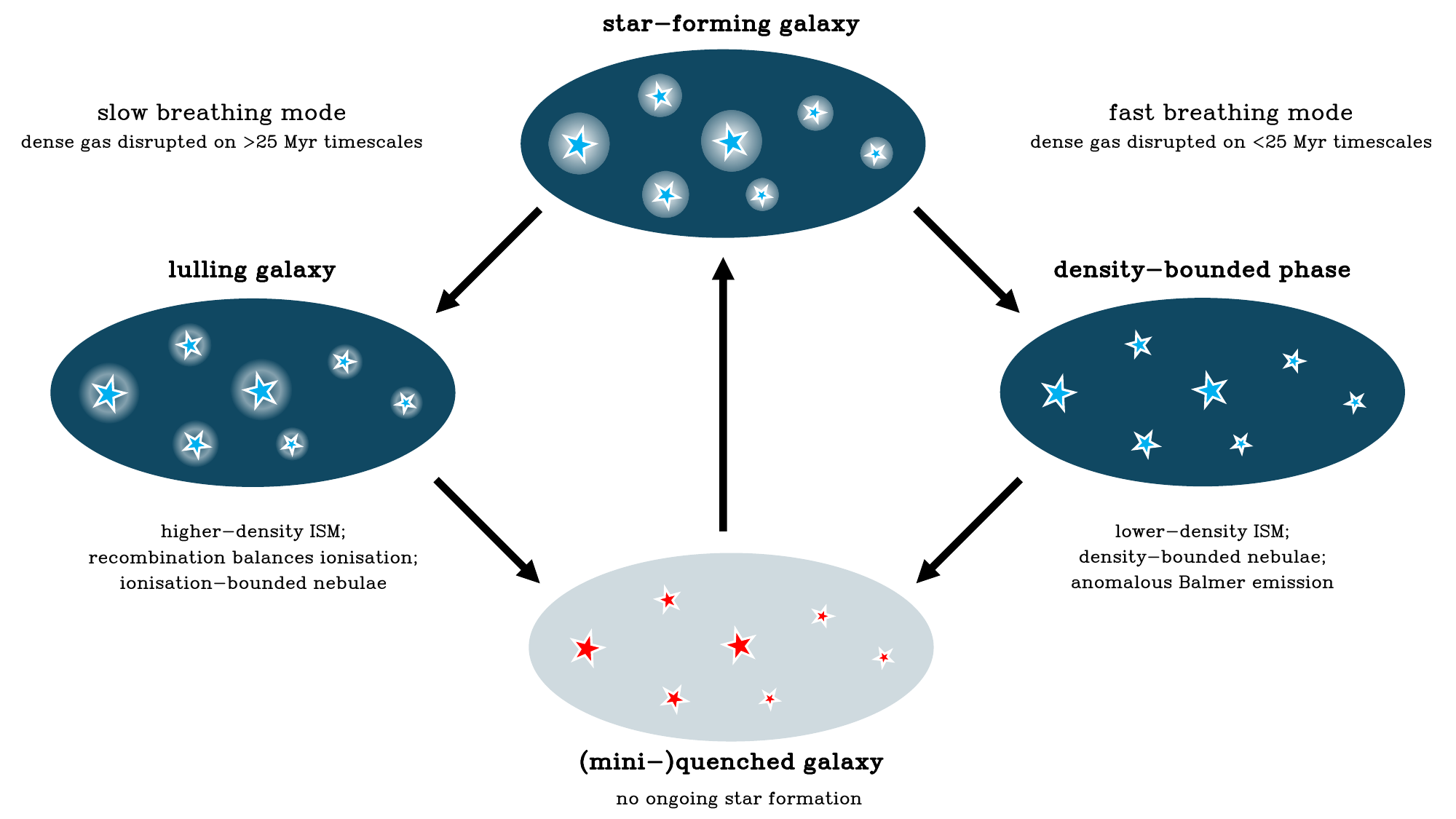}
    \caption{Schematic of a feasible formation pathway for the observed density-bounded nebulae. In the slow breathing mode, a star-forming galaxy quenches slowly relative to the lifetime of massive ionising stars ($\sim$25\,Myr), maintaining a high enough average gas density such that recombinations are sufficient to ionisation-bound the lulling galaxy. However, in the fast breathing mode, a star-forming galaxy disrupts its ISM on short timescales, such that the high-density \ion{H}{II} regions and star-formation fuel are destroyed or evacuated while there is still significant ionising emission. The remaining low-density gas is insufficient to absorb all ionising photons and the galaxy becomes a density-bounded ABE. In extreme events and for very low mass galaxies, even the low-density gas may itself be evacuated within this timeframe, and a mini-quenched galaxy will appear with little nebular emission and near-complete ionising escape.}
    \label{fig:db_scheme}
\end{figure*}

Given the robustness of our density-bounded models to changes in the stellar age of the ionising source, gas density, and ionisation parameter, it is difficult to constrain how the observed density-bounded emission arises. Nevertheless, we can consider several plausible scenarios that may contribute to movement in the Balmer line ratio diagram and explain the other spectral features of ABEs.

\citet{McClymont:2024aa} demonstrate that low-density gas in local star-forming galaxies is ionised by stars that have rapidly dissipated their \ion{H}{II} regions, and propose that a similar mechanism could operate in high-redshift galaxies. These galaxies can be dominated by intense starburst episodes and are not expected to be cloaked in such an abundance of low-density gas. Therefore, the rapid dissipation of \ion{H}{II} regions could be responsible for high-redshift mini-quenched galaxies, which show recent star formation but unexpectedly little nebular line emission \citep{Looser:2023ab,Dome:2024aa}. In this scenario, such galaxies would likely pass through an intermediate phase, transitioning from the typical ionisation-bounded nebulae in star-forming galaxies to the total ionising escape seen in mini-quenched galaxies. Galaxies in this intermediate phase would be density-bounded with low ISM density, which we discuss in this work. 

Such mini-quenching events represent the extreme end of a spectrum from bursty to smooth star formation. More commonly observed in high-redshift galaxies ($z>3$) galaxies, and lower stellar mass galaxies ($M_\star\lesssim10^{9.5}\,\Msun$) generally, is a `breathing mode' whereby starburst events catastrophically disrupt the gas in the ISM and quench star formation, but do not necessarily evacuate \textit{all} gas from the ISM within $\sim$25\,Myr \citep{El-Badry:2016aa,Tacchella:2016aa,Hopkins:2023aa, Cenci:2024aa,McClymont:2025aa,McClymont:2025ab}. During the quenching process, even if only the high-density gas is disrupted on this timescale ($\sim$25\,Myr), then the low recombination rate in the remaining low-density gas may not be sufficient to balance the ionisation, and therefore density-bounded emission may arise. In other terms, the Stromgren radius increases beyond the scale of the galaxy. Due to their density-bounded nature, galaxies in this transient phase may appear as ABEs. As the ionising radiation from the stars diminishes, the galaxy enters a quenched state, but is likely to rejuvenate again quickly. We call this process the `fast breathing mode' of star formation to indicate the timescale that the gas is disrupted. Conversely, if the high-density gas in a star-forming galaxy is disrupted on longer timescales relative to the lives of ionising massive stars, then the average gas density is always high enough, relative to the ionising output of stars, to cause the galaxy to be ionisation-bounded. We call this pathway the `slow breathing mode'.

A schematic of this formation pathway is illustrated in Fig.~\ref{fig:db_scheme}. This scenario is further motivated by the observational characteristics of ABEs, which tend to be dimmer and exhibit steeper UV slopes than their Case B consistent counterparts (see Section~\ref{sec:Continuum features and Lyman-alpha}). These traits indicate that ABEs are typically lower-mass galaxies that have recently undergone a significant starburst, leading to a much more rapid quenching event than is typical. Additionally, the presence of strong Ly$\alpha$ emission in these galaxies is consistent with a density-bounded galaxy.

Observational evidence suggests that rejuvenation following rapid quenching may also occur on short timescales (Witten et al. in prep.). We also note that this extreme (in the sense of ultra-fast on tens of Myr) quenching, or breathing mode, is distinct from the rapid quenching in literature of more massive galaxies ($M_{\star}>10^{10}\,\Msun$), which describes the long-term quenching of galaxies over timescales of several tens to a few hundred Myr \citep{Belli:2019aa,Tacchella:2022ab,DEugenio:2024ab,Park:2024aa}. This difference helps to explain the relatively more frequent ABE conditions in high-redshift galaxies.

\citet{Katz:2023ac} identified a population of density-bounded LyC leakers in the SPHINX cosmological radiation hydrodynamics simulations. These galaxies show normal ionisation parameters ($\log_{10}(q\,[\mathrm{cm\,s^{-1}}])\approx7$) and low ISM densities ($\sim$$10^{0.5}\,\textrm{cm}^{-3}$). These conditions arise at the tail end of a $\sim$100\,Myr burst of star formation which disrupts the ISM through stellar feedback, dissipating the high-density gas typical of \ion{H}{II} regions. However, the presence of a Balmer jump in our observed ABEs indicates that there cannot be a significant population of older stars relative to young ionising stars, as the Balmer break can quickly wash out the Balmer jump. Therefore, it is possible that the ISM in these galaxies is being disrupted on a shorter timescale than those seen in the SPHINX simulations. Unfortunately, directly investigating deviations from Case B in galaxy-scale simulations is currently not possible due to the lack of on-the-fly or post-processing codes with the necessary physics, although these observations can motivate further development efforts by the community \citep[e.g.][]{Smith:2022aa}.

Another possibility is a clumpy ISM characterized by small-scale, high-density clouds. Each of these high-density clouds could themselves be density-bounded, even if the galaxy as a whole is not. It may be sufficient that the brightest emission might originate from these dense, density-bounded clumps. In this scenario, although we are observing density-bounded emission, it does not necessarily imply that the entire galaxy is leaking ionising radiation.

Mergers and close encounters can both trigger star formation and disrupt the gas in galaxies. The resulting burst of star formation can lead to the observed Balmer jumps, while the disrupted gas could lead to density-bounded nebulae through processes such as gas stripping. It is also interesting to note that mergers have been associated with Ly$\alpha$ emission \citep{Witten:2024aa}, which we have shown is related to ABEs. This scenario is not necessarily independent of our preferred explanation in Fig.~\ref{fig:db_scheme}; mergers are common in the early Universe and are one potential driver of starbursts \citep{Witten:2024aa}.

The presence of an AGN could also naturally lead to a density-bounded scenario. For example, GS9422, a galaxy which shows anomalous Balmer line ratios, has already been speculated to contain an AGN \citep{Tacchella:2024aa,Li:2024aa}. However, it seems unlikely that AGN could explain all of our observed galaxies because their ionising spectra do not naturally contain the absorption features around Ly$\gamma$ which produces the turnover in the H$\alpha$/H$\beta$ versus H$\gamma$/H$\beta$ trend in stellar models (Fig.~\ref{fig:DB_Ha_Hg_vs_Ha_Hb_allmod}). Nonetheless, they may provide an alternative explanation for galaxies with the highest equivalent widths, such as GS9422.

\subsection{Complications arising from density-bounded nebulae}
\label{sec:Complications arising from density-bounded nebulae} 

The existence of density-bounded nebulae in the early Universe has several important consequences for interpreting the line emission of high-redshift galaxies. The most basic is that non-Case B values of H$\alpha$/H$\beta$ challenge a fundamental assumption used in dust correction. For galaxies with intrinsically sub-Case B H$\alpha$/H$\beta$ ratios that are partially obscured by dust attenuation, the dust correction applied would be too low. For our higher metallicity models, which have H$\alpha$/H$\beta$ values above Case B, incorrect dust corrections would be applied to reduce the H$\alpha$/H$\beta$ ratio to Case B values. It is currently difficult to understand the full extent of this impact because of the difficulty identifying ABEs that have also undergone significant dust attenuation. However, density-bounded (and potentially LyC leaking) conditions are likely correlated with lower dust attenuation, therefore we expect the impact to not be too severe.

Density-bounded nebulae may help explain the breakdown of traditional AGN diagnostics at high redshift \citep{Scholtz:2023aa, Mazzolari:2024aa}. In the cores of \ion{H}{II} regions, there is a much higher fraction of high-energy ionisation states, such as $\mathrm{O^{++}}$. The emission from density-bounded nebulae comes exclusively from the core, leading to line ratios being skewed toward high-energy ionisation states, which can mimic the effect of having a much harder ionising spectrum or higher ionisation parameter. This effect can be seen in Fig.~\ref{fig:DB_cloudy_obs_comp}, where our density-bounded models exhibit decreasing [\ion{O}{III}]/[\ion{O}{II}] with increasing cloud depth. \citet{Scholtz:2023aa} select for AGN using newly defined diagnostics for high-redshift galaxies. Interestingly, their stacked AGN sample, selected using optical emission lines, shows a Balmer decrement of 2.34, which is clearly indicative of ABEs. Either some density-bounded, star-forming galaxies have masqueraded as AGN due to the previously mentioned effects, or AGN are more likely to be density-bounded.

The escape fraction of Ly$\alpha$ is another crucial galactic property that can be severely biased by density-bounded nebulae. One common method to measure the Ly$\alpha$ escape fraction along the line of sight is to compare the observed Ly$\alpha$/H$\beta$ to the Case B Ly$\alpha$/H$\beta$ ratio. At high redshift, these measurements often yield very high Ly$\alpha$ escape fractions, above 70\% \citep{Saxena:2023aa, Witstok:2025aa}, higher than prior theoretical expectations \citep[e.g.][]{Smith:2019aa}. In density-bounded nebulae, the Ly$\alpha$/H$\beta$ ratio is skewed compared to the Case B value, and this means that the Ly$\alpha$ escape fraction can be over or underestimated. More specifically, for nebulae with very high ionising escape ($f_{\mathrm{esc}}>0.9$), the Ly$\alpha$/H$\beta$ ratio can be several times larger than the Case B ratio. However, for nebulae with more reasonable values of $f_{\mathrm{esc}}<0.9$, the Ly$\alpha$/H$\beta$ ratio can be lower by a factor of $\sim$2/3, increasing until the system becomes ionisation bounded. This issue may be even more severe because density-bounded nebulae are more likely to be Ly$\alpha$ emitters (LAEs), and are therefore likely make up a larger fraction of the LAE population than the general galaxy population.

Density-bounded nebulae can also complicate measurements of the ionising photon production efficiency, $\xi_{\mathrm{ion}}$. This quantity allows us to convert between the UV luminosity of galaxies and their intrinsic ionising photon output \citep{Stark:2016aa}. Typically, $\xi_{\mathrm{ion}}$ is estimated via the luminosity of Balmer lines, which are assumed to act as ionising photon counters \citep{Simmonds:2023aa, Simmonds:2024aa}. Density-bounded nebulae introduce two major problems for this method. Firstly, the Case B recombination assumption breaks down in density-bounded nebulae, skewing the conversion between Balmer line photons and number of ionisations. Secondly, the total amount of Balmer line photons of density-bounded nebulae is lower than their ionisation-bounded counterparts due to the high LyC escape fraction. [\ion{O}{III}] can also be a useful proxy measurement for $\xi_{\mathrm{ion}}$, this would again be skewed by density-bounded nebulae \citep{Chevallard:2018aa}. 

One consequence of this skewed measurement of $\xi_{\mathrm{ion}}$, especially combined with the complications in measuring the Ly$\alpha$ escape fraction mentioned above, is the introduction of biases in the analysis of ionised bubbles. The Ly$\alpha$ escape fraction is typically used to determine the observed size of the ionised bubble, whereas the $\xi_{\mathrm{ion}}$ and LyC $f_{\mathrm{esc}}$ is used to calculate the maximum size of ionised bubble that the observed galaxies could create \citep{Witstok:2024ab}. The comparison of the observed and maximum bubble sizes are used to infer whether the ionised bubble can be created by the observed LAE and its surrounding galaxies, or whether many unobserved, fainter galaxies are also contributing significantly. Unfortunately, density-bounded emission causes issues on both sides of this equation. The non-Case B Ly$\alpha$/H$\beta$ ratios and potentially turbulent ISM of ABEs could lead to inferred observed bubble sizes that are too large. On the other hand, the underestimation of $\xi_{\mathrm{ion}}$, would lead to maximum bubble sizes that are too small. On top of this, we expect that these galaxies have high $f_{\mathrm{esc}}$ and that they are fainter now than in their recent history, when the starburst was still ongoing, which could also lead to an underestimation of maximum bubble sizes. Given that high-redshift LAEs have the similar characteristics as ABEs \citep{Witstok:2025aa}, the potential issues arising for density-bounded emission should be given consideration.

\subsection{Alternative explanations for anomalous Balmer line ratios}
\label{sec:Alternative explanations for anomalous Balmer line ratios} 

We do not claim to completely rule out other non-Case B scenarios, as it is possible that other combinations of Balmer optical depths, dust, and geometry could potentially reproduce the observed line ratios. However, we find density-bounded nebulae to be a particularly compelling explanation for ABEs in comparison to previous work \citep[e.g.][]{Yanagisawa:2024ac,Scarlata:2024aa}. In fact, density-bounded nebulae are expected to exist in the early Universe, as leaking radiation from galaxies is responsible for driving cosmic reionisation. Our density-bounded models appear to be consistent with all available data for Balmer line ratios and, due to the use of stellar SEDs with varying metallicity, we find there is no need for fine-tuning the depth of models. Density-bounded nebulae are also naturally associated with high \ion{O}{II}/\ion{O}{III} ratios, which we have shown are found in ABEs and which have previously been associated with LyC leakers \citep{Nakajima:2014aa,Nakajima:2020aa}.

Additionally, our models predict parameters with H$\beta$ equivalent widths that are consistent with the data. However, we caution against using the equivalent width as measured from geometrically simple \texttt{CLOUDY} models to rule out parameter space for density-bounded nebulae. Galaxies are complex systems with non-trivial distributions of dust and gas. That is even more true for density-bounded systems, which by definition have unusual distributions of their gas. Fig.~\ref{fig:DB_Ha_Hg_vs_Ha_Hb_allmod} shows that observed galaxies with low H$\alpha$/H$\beta$ are consistent with, if not require, some degree of dust attenuation. It is not unreasonable to posit some geometric effects leading to the stellar continuum being more heavily attenuated than the emission lines in these density-bounded systems.

One alternative explanation for ABEs, which is closely related to our density-bounded models, are Case D nebulae. These nebulae have such significant line pumping that they show sub-Case B H$\alpha$/H$\beta$ even when ionisation-bounded. An example of this is the model presented in Fig.~\ref{fig:D_FESC} of 10 Myr stars ionising turbulent gas (140 $\mathrm{km\,s^{-1}}$). However, we find this scenario less convincing than density-bounded models because the gas would have to be very turbulent, or the continuum very bright around the Lyman lines, in order to explain the more extreme deviations from Case B.

\section{Conclusions}
\label{sec:Conclusions}

We analysed galaxies in the JADES survey and identified 52 galaxies at $z>2$ and 14 galaxies at $z>5.3$ with Balmer decrements that are more than $1\sigma$ below the expected Case B values. We also found 12 galaxies at $z>5.3$ with H$\gamma$/H$\beta$ ratios more than $1\sigma$ above Case B values (Fig.~\ref{fig:Ha_Hb_caseB}). The line ratios in these anomalous Balmer emitters (ABEs) cannot be explained by dust attenuation alone, indicating a departure from standard Case B physics (Fig.~\ref{fig:Ha_Hg_vs_Ha_Hb_caseB}).

We created models of density-bounded nebulae using the photoionisation code \texttt{CLOUDY}. These models produce anomalous Balmer line ratios that vary depending on the depth of the nebula (Fig.~\ref{fig:DB_radial_plot}). For small cloud depths, the gas is optically thin to Lyman lines and continuum pumping of Lyman lines is important; this is equivalent to Case C. As the cloud depth increases, the cloud becomes optically thick to Lyman lines and continuum pumping of Lyman lines becomes unimportant; this corresponds to Case B.

The H$\alpha$/H$\beta$ versus H$\gamma$/H$\beta$ trends of our density-bounded nebulae models are robust against changes in the stellar age of the ionising source, gas density, and ionisation parameter (Fig.~\ref{fig:DB_Ha_Hg_vs_Ha_Hb_allmod}). However, varying metallicity drives a turnover in the relationship, allowing our models to account for all observed galaxies.

The turnover is caused by absorption features around Ly$\gamma$ in stellar SEDs, which increase with higher metallicity (Fig.~\ref{fig:bpass_plot}); specifically, the absence of these absorption features makes low-metallicity stars more efficient than high-metallicity stars at pumping electrons to the $n=4$ state of hydrogen through the Ly$\gamma$ transition, causing an excess of H$\beta$ emission. Density-bounded nebulae can mimic the line ratios of models with harder intrinsic ionising SEDs or with higher ionisation parameter (Fig.~\ref{fig:DB_cloudy_obs_comp}).

There are several scenarios that may lead to the formation of density-bounded nebulae in the early Universe. One promising interpretation is that density-bounded emission arises as a transient phase during a fast breathing mode of star formation which occurs in lower-mass galaxies, and we show this pathway in Fig.~\ref{fig:db_scheme}. If the high-density gas in a star-forming galaxy is disrupted on a timescale comparable to the lifetime of massive ionising stars ($\sim$25\, Myr), then the remaining low-density gas will be insufficient to ionisation-bound the ionising radiation from the young stars. This causes a density-bounded galaxy which would be observed as an ABE. The most extreme of these events may also eject the low-density gas within this timescale, leading to a mini-quenched galaxy with little nebular emission and total ionising escape.

Further investigation of anomalous Balmer line ratios may pay dividends for our understanding of feedback processes that regulate the ISM in the earliest galaxies. While there are many observational avenues to explore, the easing of Case B assumptions in galaxy-scale theoretical work is likely to be particularly fruitful.

\section*{Acknowledgements}

WM thanks the Science and Technology Facilities Council (STFC) Center for Doctoral Training (CDT) in Data intensive Science at the University of Cambridge (STFC grant number 2742968) for a PhD studentship. ST acknowledges support by the Royal Society Research Grant G125142. RM, WM, FDE, XJ, and JW acknowledge support by the Science and Technology Facilities Council (STFC), by the ERC through Advanced Grant 695671 ``QUENCH'', and by the UKRI Frontier Research grant RISEandFALL. RM also acknowledges funding from a research professorship from the Royal Society. CW thanks the Science and Technology Facilities Council (STFC) for a PhD studentship, funded by UKRI grant 2602262.

%%%%%%%%%%%%%%%%%%%%%%%%%%%%%%%%%%%%%%%%%%%%%%%%%%
\section*{Data Availability}

NIRSpec spectra are publicly available on MAST (\url{https://archive.stsci.edu/hlsp/jades}), with \doi{10.17909/8tdj-8n28}, \doi{10.17909/z2gw-mk31}, and \doi{10.17909/fsc4-dt61}.



\bibliographystyle{mnras}
\bibliography{db_mcclymont} % if your bibtex file is called example.bib

%\begin{thebibliography}{99}
%\end{thebibliography}

%%%%%%%%%%%%%%%%%%%%%%%%%%%%%%%%%%%%%%%%%%%%%%%%%%

%%%%%%%%%%%%%%%%% APPENDICES %%%%%%%%%%%%%%%%%%%%%

\appendix

\section{Impact of stellar age}
\label{sec:Impact of stellar age}

\begin{figure}
    \centering
	\includegraphics[width=\columnwidth]{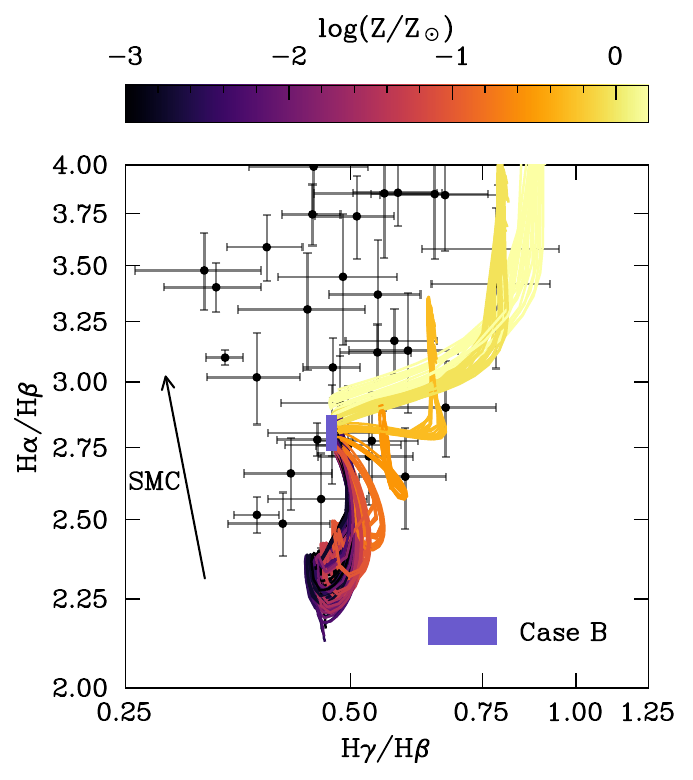}
    \caption{The same as Fig.~\ref{fig:DB_cloudy_obs_comp} except only showing models with 3\,Myr stars. }
    \label{fig:DB_Ha_Hg_vs_Ha_Hb_3myr}
\end{figure}

\begin{figure}
    \centering
	\includegraphics[width=\columnwidth]{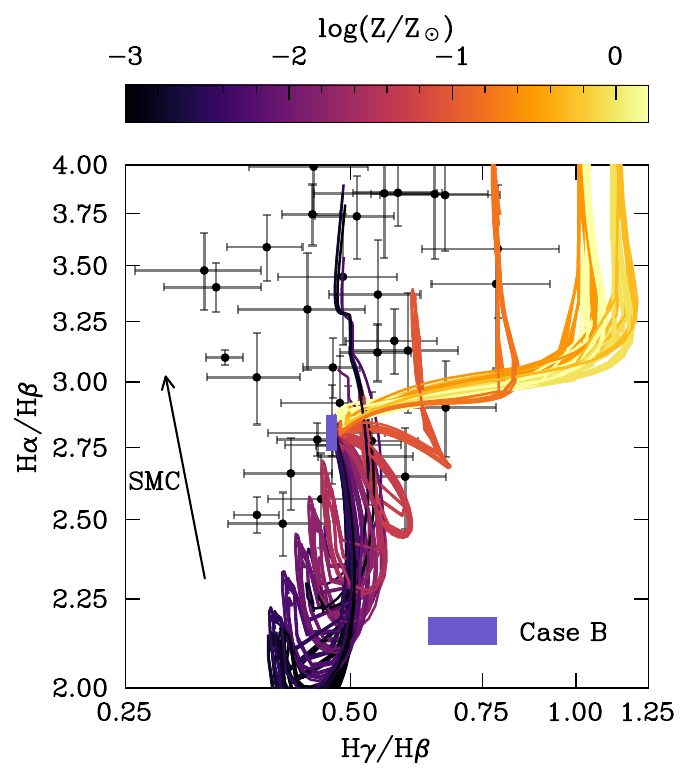}
    \caption{The same as Fig.~\ref{fig:DB_cloudy_obs_comp} except only showing models with 10\,Myr stars. }
    \label{fig:DB_Ha_Hg_vs_Ha_Hb_10myr}
\end{figure}

In Fig.~\ref{fig:DB_Ha_Hg_vs_Ha_Hb_3myr} and Fig.~\ref{fig:DB_Ha_Hg_vs_Ha_Hb_10myr} we show the Balmer line ratios seen in models as tracks following increasing cloud depth (as in Fig.~\ref{fig:DB_Ha_Hg_vs_Ha_Hb_allmod}) for only 3\,Myr and 10\,Myr stars respectively. We can see that both ages have the same general trends as a function of metallicity. However, it is clear that older stars show generally more extreme offsets from Case B. The relatively small spread at a fixed stellar age and metallicity shows that electron density and ionisation parameter has a generally small impact.

%%%%%%%%%%%%%%%%%%%%%%%%%%%%%%%%%%%%%%%%%%%%%%%%%%

% Don't change these lines
\bsp	% typesetting comment
\label{lastpage}
\end{document}